\documentclass[10pt,conference]{IEEEtran}

\usepackage{cite}
\usepackage{amsmath,amssymb,amsthm}
\usepackage{graphicx}                        
\usepackage{float}
\usepackage{physics}
\usepackage[table]{xcolor}
\usepackage[T1]{fontenc}
\usepackage[caption=false,font=footnotesize]{subfig}

\allowdisplaybreaks

\theoremstyle{remark}
\newtheorem*{remark*}{Remark}
\newtheorem*{note*}{Note}

\usepackage{tikz}

\begin{document}

\title{Fault Tolerant Decoding of QLDPC-GKP Codes \\ with Circuit Level Soft Information}

\author{
    \IEEEauthorblockN{Shantom K. Borah, Asit K. Pradhan, Nithin Raveendran, Michele Pacenti, and Bane Vasi\'{c}}
    \IEEEauthorblockA{\textit{Dept. of Electrical and Computer Engineering, The University of Arizona, Tucson, AZ, 85721 USA}\\ 
    Email: \{shantomborah, asitpradhan, nithin, mpacenti\}@arizona.edu, vasic@ece.arizona.edu}}

\maketitle

\begin{abstract}
Concatenated bosonic-stabilizer codes have recently gained prominence as promising candidates for achieving low-overhead fault-tolerant quantum computing in the long term. In such a system, analog information obtained from the syndrome measurements of an inner bosonic code is used to inform decoding for an outer code layer consisting of a discrete-variable stabilizer code such as a surface code. The use of Quantum Low Density Parity Check (QLDPC) codes as an outer code is of particular interest in view of the significantly higher encoding rates offered by these code families, leading to a further reduction in the overhead requirements for large-scale quantum computing. Recent works have investigated the performance of QLDPC-GKP codes in detail, and indeed, the use of analog information obtained from the inner code significantly boosts the decoder performance. However, the noise models assumed in these works are typically limited to the depolarizing or phenomenological noise models. In this paper, we investigate the performance of QLDPC-GKP concatenated codes under circuit-level noise, based on a noise model introduced by Noh et al. in relation to the surface-GKP code. To demonstrate the performance boost provided by the use of analog information, we investigate three different scenarios, namely (a) decoding performance in the absence of soft information, (b) decoding performance with precomputed probabilities of error but without real-time soft information, and (c) decoding performance with real-time soft information obtained from round-to-round decoding of the inner GKP code. Results show minimal performance improvement between (a) and (b), but a significant boost in (c), indicating that real-time soft information is critical for concatenated decoding under circuit-level noise. We also investigate the effect of using measurement schedules of varying depths and show that using a schedule with minimum depth is critical for obtaining reliable soft information from the inner code.
\end{abstract}

\IEEEpeerreviewmaketitle

\section{Introduction}
Classical error correction systems often utilize the idea of coded modulation, wherein the encoding process is integrated with the modulation scheme, allowing the error correction code to achieve lower error rates by exploiting the characteristics of the communication channel. In the context of recent research in the field of quantum error correction, this has taken the form of concatenated bosonic-stabilizer codes where a bosonic code such as the Gottesman-Kitaev-Preskill (GKP) code plays the role of a modulation scheme, with soft information obtained from the GKP syndrome measurements being utilized by an outer discrete-variable stabilizer code for improving decoding performance. In such a system, a bosonic code such as a GKP code \cite{gottesman2001encoding,grimsmo2021quantum} is used to embed a qubit into the Hilbert space of a quantum harmonic oscillator and the logical qubits of this bosonic code are subsequently used as the physical qubits for an outer discrete-variable code. Several recent works in the quantum error correction space have focused on such concatenated schemes with varying choices of inner bosonic codes and outer stabilizer codes \cite{zhang2021quantum, zhang2023concatenation, raveendran2022finite, berent2024analog, li2024correcting, noh2020fault, noh2022low, vuillot1a2018quantum} and have successfully demonstrated significant improvements in error rates via the use of analog \emph{soft information} obtained from the syndrome measurement rounds of the inner bosonic code. From a hardware perspective, several prominent hardware platforms for quantum computing, including superconducting qubits \cite{campagne2020quantum, blais2021circuit} as well as photonic qubits \cite{tzitrin2020progress} are known to support an infinite-dimensional Hilbert space that is isomorphic to that of a quantum harmonic oscillator, which positions these concatenated architectures as a compelling candidate for low-overhead fault tolerance in the long term. 

Concatenated schemes with the GKP code as the inner code and a surface code as the outer code were first investigated by Noh et al. in \cite{noh2020fault}, where a soft Minimum Weight Perfect Matching (MWPM) decoder, with edge weights renormalized in accordance with soft information from the inner GKP code, was used to improve decoding performance for the outer surface code. Subsequently, in \cite{noh2022low}, the authors expanded their noise model to account for noise correlations due to CNOT gates in the syndrome measurement circuit for the outer code and developed a variant of the soft MWPM decoder that is able to handle the effects of these correlations. Finally, \cite{lin2023closest} demonstrated yet another significant improvement in the decoder performance through the use of an optimal closest point decoder that operates on the multimode GKP lattice of the surface-GKP code. Error correction architectures involving the concatenation of the GKP code with several other topological codes have also been studied extensively. These include, among others, the toric-GKP code \cite{vuillot1a2018quantum}, the color-GKP code \cite{zhang2021quantum}, and the XZZX surface-GKP code \cite{zhang2023concatenation}.

While these topological quantum codes \cite{dennis2002topological} have several promising properties in the context of fault tolerance \cite{fowler2012surface, litinski2019game}, they suffer from the problem of a vanishing encoding rate \cite{roffe2019quantum}. Indeed, the Bravyi-Terhal Theorem \cite{bravyi2009no} states that any 2D-local topological quantum stabilizer code, where each stabilizer generator acts on a constant number of qubits, has a code distance that scales at most logarithmically with the system size. This leads to a rather prohibitive increase in the overhead requirements for large-scale fault tolerant quantum computing. To counter this, several classes of quantum LDPC codes \cite{breuckmann2021quantum} have been proposed that provide a better scaling of minimum distance and encoding rate \cite{macwilliams1977theory} with respect to an increase in the system size at the cost of requiring long range non-local qubit connectivity. Starting with the hypergraph product construction \cite{tillich2013quantum} by Tillich \& Z\'{e}mor, several constructions of QLDPC codes have since been proposed that achieve an increasingly higher minimum distance scaling with the code length, while maintaining a finite rate. These include fiber bundle codes \cite{hastings2021fiber}, lifted product codes \cite{panteleev2021quantum}, and balanced product codes \cite{breuckmann2021balanced}.

The performance of concatenated QLDPC-GKP codes under the depolarizing noise model \cite{nielsen2010quantum} was first investigated by Raveendran et al. in \cite{raveendran2022finite}. Subsequently, the performance of these codes under the phenomenological noise was investigated by Berent et al. in \cite{berent2024analog}. In this work, we forge the first steps in characterizing the performance of concatenated QLDPC-GKP codes under the circuit-level noise model. Specifically, we adopt the noise model described in \cite{noh2022low} and operate under the assumption that the major source of noise in the system is the finite squeezing of the ancilla qubits used in the GKP code. Under this simplifying assumption, we characterize the performance of two QLDPC code variants in three separate scenarios, namely, (a) decoding in the absence of soft information, (b) decoding with precomputed reliability information, and (c) decoding with real-time soft information. Our observations indicate that a significant improvement in the frame error rate may only be obtained in the third case, indicating that real-time soft information plays a critical role for decoding under circuit-level noise. We also investigate the effect of idling faults incurred by using measurement circuits of varying depth. In this context, our observations indicate that the reliability of the soft information obtained from the inner code depends strongly on the depth of the measurement circuit. Hence, an optimized measurement schedule with minimum depth is essential for the proper utilization of soft information from the inner code by the outer decoder. The rest of this paper is organized as follows. We begin by reviewing some background material in relation to the GKP code in Sec. \ref{sec:gkp_code}. We also outline the assumptions involved in the noise model that we shall use to capture the effect of noisy GKP error correction. Subsequently, we provide a description of the circuit-level noise model for the outer QLDPC code and the associated decoding process in Sec. \ref{sec:qldpc_code}. Finally, we present the results of our Monte Carlo simulations with the concatenated QLDPC-GKP code under circuit-level noise in Sec. \ref{sec:results} and conclude with a summary and discussion of future research directions in Sec. \ref{sec:conclusion}.

\section{The GKP Code}
\label{sec:gkp_code}
We will begin in this section, with a brief introduction to the Gottesman-Kitaev-Preskill (GKP) code \cite{gottesman2001encoding, grimsmo2021quantum}, which will play the role of the inner code in a concatenated QLDPC-GKP architecture. Subsequently, we shall provide a detailed description of the circuit-level noise model for a concatenated QLDPC-GKP scheme. Specifically, we will adopt the noise model developed in \cite{noh2022low} and describe the fault mechanisms through which the finite squeezing of the GKP ancilla qubits can translate into circuit-level noise for the outer QLDPC code.

\subsection{GKP Error Correction}
\label{sec:gkp_qec}
A bosonic code is a type of error correction code that embeds a set of discrete-variable quantum systems into a collection of continuous-variable quantum systems. Typically, the physical layer of such a code corresponds to one or more oscillator modes \cite{blais2021circuit}, with each mode representing the Hilbert space of a quantum harmonic oscillator. The logical space of such a code is typically that of a finite-dimensional discrete-variable quantum system such as a qubit or qudit \cite{borah2024non}. One of the simplest variants of bosonic codes is the GKP code, which embeds a single qubit into a single oscillator mode. Formally, the GKP code is a stabilizer code with the following displacement operators as its stabilizers.
\begin{equation}
    \hat{S}_q = \exp{(i2\sqrt{\pi}\hat{q})}; \qquad \hat{S}_p = \exp{(-i2\sqrt{\pi}\hat{p})},
    \label{eq:gkp_stabilizers}
\end{equation}
where $\hat{q}$ is the position operator and $\hat{p}$ is the momentum operator in the phase space of the quantum harmonic oscillator \cite{sakurai202modern}. In other words, the GKP code-space is the common eigenspace of the two operators above with the eigenvalue +1. Alternatively, the GKP code-space can also be described in terms of the logical states with respect to the $X$ and $Z$ bases as follows.
\begin{equation}
\begin{split}
    \ket{0}\; &= \sum_{n \in 2\mathbb{Z}} \ket{\hat{q}=n\sqrt{\pi}}; \\
    \ket{+} &= \sum_{n \in 2\mathbb{Z}} \ket{\hat{p}=n\sqrt{\pi}};
\end{split}\qquad
\begin{split}
    \ket{1}\; &= \sum_{n \in 2\mathbb{Z}+1} \ket{\hat{q}=n\sqrt{\pi}}; \\
    \ket{-} &= \sum_{n \in 2\mathbb{Z}+1} \ket{\hat{p}=n\sqrt{\pi}},
\end{split}
\label{eq:gkp_codewords}
\end{equation}
where $\ket{\hat{q}=q_0}$ and $\ket{\hat{p}=p_0}$ represent eigenvectors of the $\hat{q}$ and $\hat{p}$ operators with eigenvalue $q_0$ and $p_0$, respectively. Hence, each logical state in the above equation is a comb of Dirac-delta wave functions (either in the position or momentum basis) with the $\ket{0}$ state (similarly $\ket{+}$) having peaks centered at even multiples of $\sqrt{\pi}$ and the $\ket{1}$ state (similarly $\ket{-}$) having peaks centered at odd multiples of $\sqrt{\pi}$.

In practice, these Dirac-delta combs are infinite energy states and cannot be physically implemented on any realistic hardware. To describe physically implementable finite-energy GKP states, we will make use of the twirling approximation introduced by Noh et al. in \cite{noh2020fault}, wherein finite-energy GKP states can be understood as a combination of ideal GKP states followed by a set of Gaussian shift errors, with a noise variance of $\sigma^2$, in both the position and momentum quadratures. Following standard conventions \cite{noh2022low}, we define the GKP squeezing in decibels as follows:
\begin{equation}
    \sigma_{\text{GKP}}^{(\text{dB})} = 10 \log_{10}\left(\frac{1}{2\sigma^2}\right).
    \label{eq:squeezing_param}
\end{equation}

To simulate the effect of GKP error correction, we sample random Gaussian shifts $\eta_q, \eta_p$ from the normal distribution $\mathcal{N}(0, \sigma^2)$. We declare a logical $X$ error if $\eta_q$ is closer to the nearest odd multiple of $\sqrt{\pi}$ than to the nearest even multiple of $\sqrt{\pi}$. In other words, a logical $X$ error occurs if,
\begin{equation}
    \eta_q \in \bigcup_{n \in \mathbb{Z}}\left[\left(\frac{4n + 1}{2}\right)\sqrt{\pi}, \left(\frac{4n + 3}{2}\right)\sqrt{\pi}\right].
    \label{eq:gkp_error_region}
\end{equation}
Similarly, a logical $Z$ error occurs if the shift $\eta_p$ lies in the above region. Hence, for a given noise variance of $\sigma^2$, the probability of a logical error (of either $X$ or $Z$ type) may be evaluated as 
\begin{equation}
    P_{\sigma^2}(\text{error}) = \frac{1}{\sqrt{2\pi\sigma^2}}\sum_{n \in \mathbb{Z}}\int_{\left(\frac{4n + 1}{2}\right)\sqrt{\pi}}^{\left(\frac{4n + 3}{2}\right)\sqrt{\pi}} \exp\left(-\frac{x^2}{2\sigma^2}\right) dx.
    \label{eq:gkp_prior}
\end{equation}
An important advantage of GKP error correction is the generation of analog soft information about the GKP data qubits during the measurement of the GKP stabilizers. This soft information essentially corresponds to a measurement of the analog shift error $\eta_q$ (or $\eta_p$) modulo $\sqrt{\pi}$. Given this measurement of $\eta$, we may thus obtain a more precise estimate of the probability of logical error by using the following equation \cite{raveendran2022finite}.
\begin{equation}
    P_{\sigma^2}(\text{error}|\eta) = \frac{\sum_{n \in \mathbb{Z}} \exp{\left(\frac{-(\eta - (2n+1)\sqrt{\pi})^2}{2 \sigma^2}\right)}}{\sum_{n \in \mathbb{Z}} \exp{\left(\frac{-(\eta - n\sqrt{\pi})^2}{2 \sigma^2}\right)}}.
    \label{eq:gkp_posterior}
\end{equation}

In general, the GKP code features two well-known protocols to correct for small shift errors in the position and momentum quadratures for the data qubits, namely, the Steane error correction protocol and the teleportation-based error correction protocol \cite{grimsmo2021quantum}. In \cite{noh2022low}, it was shown that the teleportation-based protocol leads to a lower variance in the propagation of finite-squeezing errors from the ancilla to the data qubits. Following the same reasoning, we will assume the use of the teleportation-based protocol in this work, although we note that the same methods also apply to the Steane error correction protocol. In \cite{noh2022low}, the authors also introduced a detailed noise model for the concatenated surface-GKP code, in which the main source of noise is the finite energy of practical GKP states as described above. In this work, we make a similar assumption and adapt elements of the same noise model for the circuit-level decoding of QLDPC codes. We will now provide a brief overview of this noise model.
\begin{figure*}
    \centering
    \hspace{2.5em}\subfloat[Noise model for GKP error correction.]{\includegraphics[width=0.45\textwidth]{
        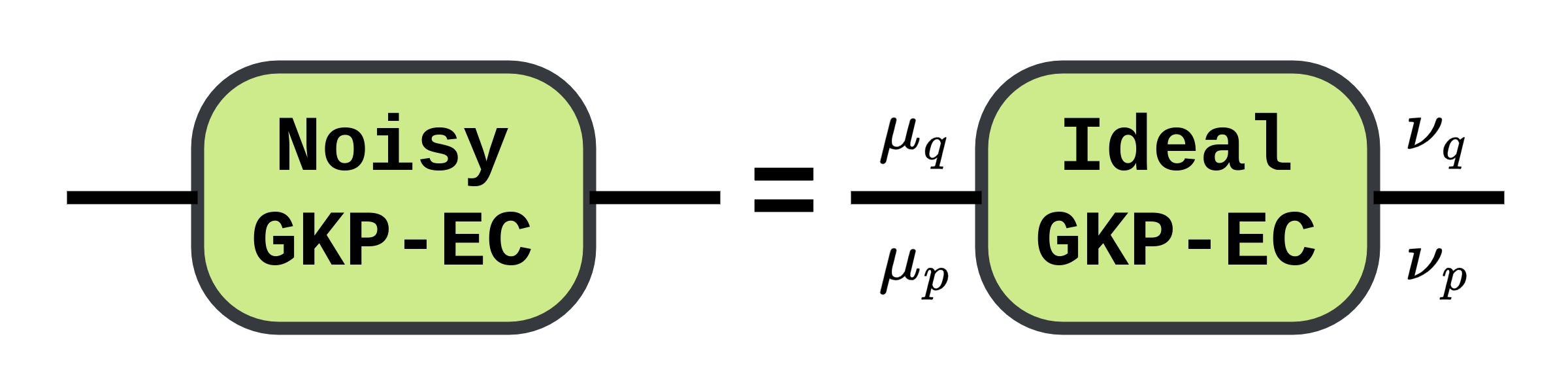}
        \label{fig:gkp_noise}}\hfill
    \subfloat[Noise model for faulty state preparation.]{
        \includegraphics[width=0.45\textwidth]{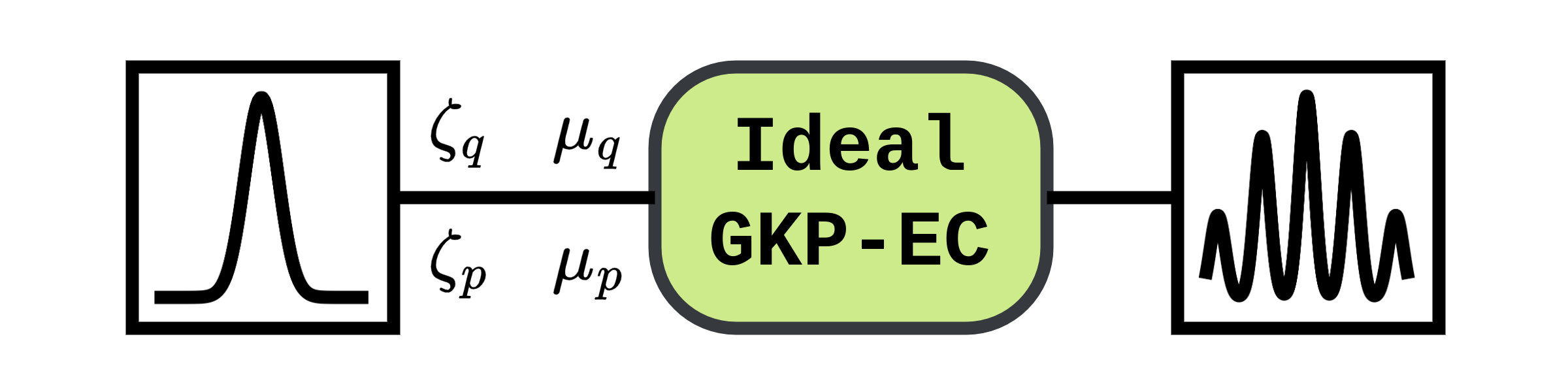}
        \label{fig:gkp_prep}}\qquad\phantom{0}\\
    \raisebox{0.78\height}{ % Adjust the height to align properly
    \hspace{1.9em}\begin{minipage}{0.45\textwidth}
        \subfloat[Noise model for idling fault.]{
            \includegraphics[width=\textwidth]{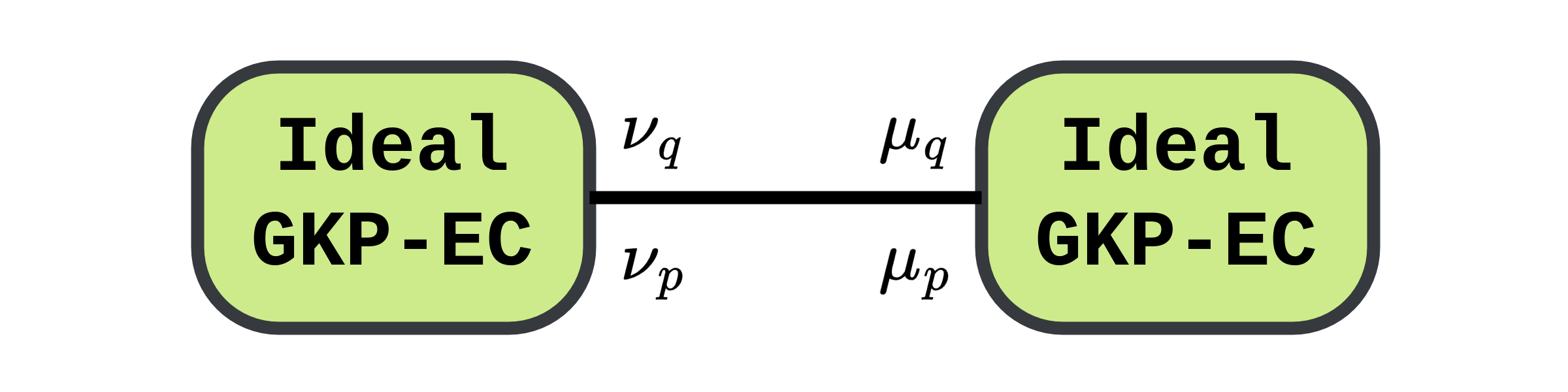}
            \label{fig:gkp_idle}}\\
        \subfloat[Noise model for faulty measurement.]{
            \includegraphics[width=\textwidth]{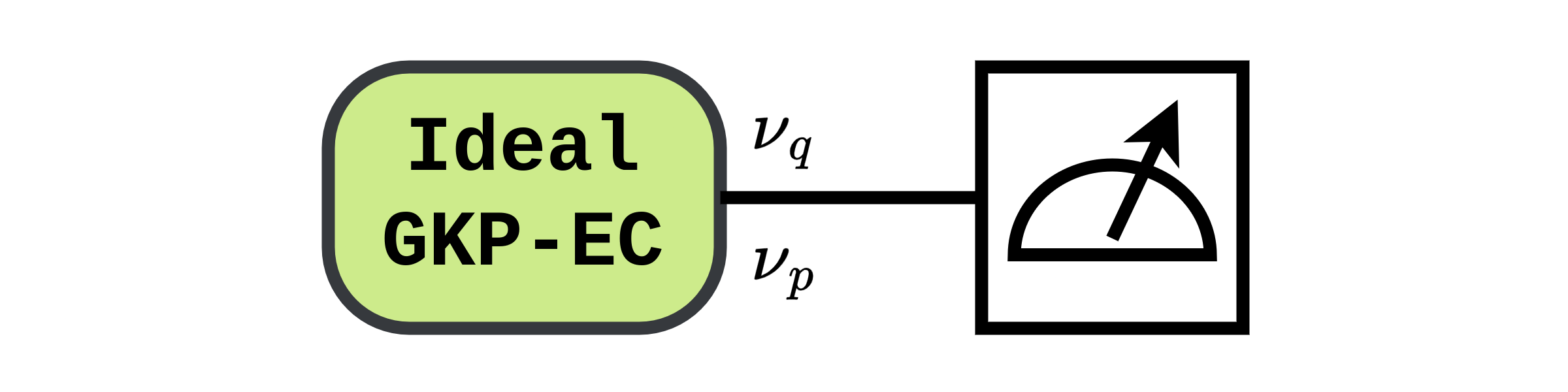}
            \label{fig:gkp_meas}}    
    \end{minipage}
    }\hfill
    \subfloat[Noise model for CNOT gate faults.]{
        \includegraphics[width=0.45\textwidth]{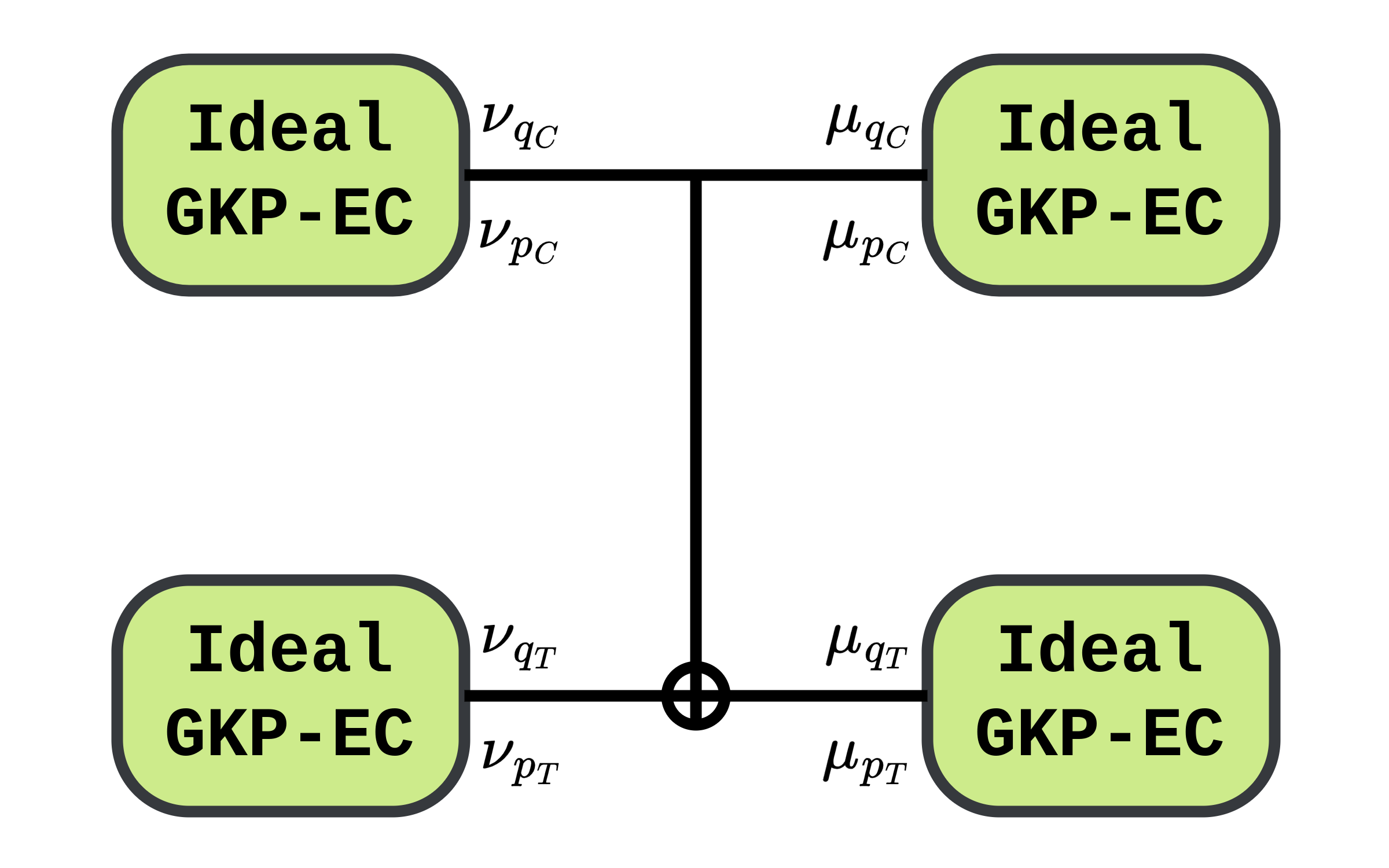}
        \label{fig:gkp_gate}}
    \caption{Noise model for noisy GKP error correction with finitely squeezed GKP ancilla qubits. (a) The noisy GKP block can be modeled as an ideal GKP block preceded by a set of prior shifts and followed by a set of posterior shifts. Under this assumption, error models are developed for four different scenarios, namely, (b) state preparation, (c) idling, (d) measurement and (e) the action of CNOT gates (to be used for syndrome measurements in the outer code). We use the subscripts $q$ and $p$ for indicating shifts in the position and momentum quadratures respectively. For the case of CNOT gate faults, we use subscripts $q_C, p_C$ for shifts on the control qubit and $q_T, p_T$ for shifts on the target qubit.}
    \label{fig:gkp_faults}
\end{figure*}

\subsection{The Noisy GKP Protocol}
\label{sec:gkp_noise}
As mentioned in the previous section, the finite squeezing of the GKP ancilla qubits is a major source of noise in the GKP error correction protocol, which can be approximately modeled as a Gaussian shift error drawn from a normal distribution with a certain GKP squeezing parameter as defined in Eq. \ref{eq:squeezing_param}. It was shown in \cite{noh2022low} that the effect of these finite squeezing errors can be succinctly captured using the model shown in Fig. \ref{fig:gkp_noise}. Specifically, a noisy GKP error correction block may be modeled as an ideal GKP error correction block that is, 
\begin{itemize}
    \item[(1)] preceded by a set of shifts $\mu_q, \mu_p \in \mathcal{N}(0, \sigma_{\text{GKP}}^2)$ in the position and momentum quadratures, respectively.
    \item[(2)] followed by a set of shifts $\nu_q, \nu_p \in \mathcal{N}(0, \sigma_{\text{GKP}}^2)$ in the position and momentum quadratures, respectively.
\end{itemize}
We shall refer to the shifts $\mu_q, \mu_p$ as the \emph{prior shifts} and the shifts $\nu_q, \nu_p$ as the \emph{posterior shifts} of the GKP error correction block. Note that all of these four shifts correspond to independent Gaussian random variables drawn from a normal distribution with a variance of $\sigma_{\text{GKP}}^2$, where $\sigma_{\text{GKP}}$ is the GKP squeezing parameter.

Having defined the noise model for a single GKP block, we will now consider the interaction of such noisy blocks in several scenarios of interest for a concatenated QLDPC-GKP architecture. Fig. \ref{fig:gkp_faults} shows four such scenarios, namely, idling, state preparation, measurement and CNOT gates. For each of these four scenarios, our aim would be to understand how the continuous-variable Gaussian shifts of Fig. \ref{fig:gkp_faults} lead to discrete-variable Pauli faults for the outer code.

\subsubsection{Idling Faults}
An idling fault occurs when two rounds of GKP Error Correction (henceforth, GKP-EC) are applied to a qubit with no gate operation on the qubit in between these two rounds, as shown in Fig. \ref{fig:gkp_idle}. In this case, the effective shifts entering the latter GKP-EC block are given by,
\begin{equation}
    \eta_q = \mu_q+\nu_q; \qquad \eta_p = \mu_p+\nu_p.
\end{equation}
Since the shifts $\mu_q, \mu_p, \nu_q, \nu_p$, are all independent the effective noise is a set of Gaussian shifts $\eta_q, \eta_p$ with variance $2\sigma_{\text{GKP}}^2$. Hence, to sample an idling fault, we simply sample a shift from $\mathcal{N}(0, 2\sigma_{\text{GKP}}^2)$ and declare a fault if it falls within the region shown in Eq. \ref{eq:gkp_error_region}. The fault occurrence probability can then be evaluated by Eq. \ref{eq:gkp_prior} with $\sigma^2 = 2\sigma_{\text{GKP}}^2$. Subsequently, a sample measurement outcome of $\eta$ obtained from the GKP-EC blocks gives a more accurate estimate of the fault occurrence probability via Eq. \ref{eq:gkp_posterior}.

\subsubsection{Preparation Faults}
For state preparation in the $\ket{0}$ state, we assume that a squeezed state $\ket{\psi}$ is prepared in the $q$ quadrature, with a squeezing parameter corresponding to a variance of $\sigma_{\text{GKP}}^2$, and subsequently, this is projected into the GKP $\ket{0}$ state via a GKP-EC block. This is shown in Fig. \ref{fig:gkp_prep}. We have two sources of noise, a set of Gaussian shifts $\zeta_q, \zeta_p$ with variance $\sigma_{\text{GKP}}^2$ due to the finite squeezing of the state $\ket{\psi}$ and the prior shifts $\mu_q, \mu_p$ preceding the GKP-EC block. The effective shifts entering the latter GKP-EC block are given by,
\begin{equation}
    \eta_q = \zeta_q+\mu_q; \qquad \eta_p = \zeta_p+\mu_p.
\end{equation}
The effective variance is then again equal to $\sigma^2 = 2\sigma_{\text{GKP}}^2$ and the behavior of a preparation fault is hence, identical to that of an idling fault.

\subsubsection{Measurement Faults}
The case of measurement faults is shown in Fig. \ref{fig:gkp_meas}. In this case, there is only one source of error, i.e., the posterior shifts $\nu_q, \nu_p$ following the last GKP block prior to measurement. The effective shifts are
\begin{equation}
    \eta_q = \nu_q; \qquad \eta_p = \nu_p.
\end{equation}
Hence, we have an effective variance of $\sigma_{\text{GKP}}^2$ for this type of fault. Fault sampling and the evaluation of fault occurrence probabilities can then be done similarly using Eq. \ref{eq:gkp_error_region}, Eq. \ref{eq:gkp_prior}, and Eq. \ref{eq:gkp_posterior} with $\sigma^2 = \sigma_{\text{GKP}}^2$.

\subsubsection{Gate Faults}
Gate faults occur when a CNOT gate is applied to a set of two GKP qubits between two rounds of GKP-EC, as shown in Fig. \ref{fig:gkp_gate}. In this case, there are eight independent Gaussian shifts, each with a variance of $\sigma_{\text{GKP}}^2$. Note that each of these shifts are essentially displacement operators \cite{sakurai202modern} acting on the physical Hilbert spaces of the GKP qubits and that the CNOT gate is physically implemented as a CSUM gate \cite{raveendran2022finite} between the qubits. We can thus, conjugate the posterior shifts $\nu_{q_C}, \nu_{p_C}, \nu_{q_T}, \nu_{p_T}$ through the CSUM gate and add these to the prior shifts $\mu_{q_C}, \mu_{p_C}, \mu_{q_T}, \mu_{p_T}$ to get the following effective shifts on the control and target qubits.
\begin{equation}
\begin{split}
    \eta_{q_C} &= \nu_{q_C} + \mu_{q_C} \\
    \eta_{q_T} &= \nu_{q_T} + \nu_{q_C} + \mu_{q_T} \\
    \eta_{p_C} &= \nu_{p_C} - \nu_{p_T} + \mu_{p_C} \\
    \eta_{p_T} &= \nu_{p_T} + \mu_{p_T}
    \label{eq:gkp_cnot}
\end{split}
\end{equation}
Note that the effective shifts of the control and target qubits are now correlated. Thus, we can describe the probability distributions for the effective shifts using the following joint Gaussian distributions.
\begin{equation}
\begin{split}
    P(\eta_{q_C}, \eta_{q_T}) &= \frac{1}{\mathcal{N}_0} \exp \left[ -\frac{3\eta_{q_C}^2 + 2\eta_{q_T}^2 - 2\eta_{q_C}\eta_{q_T}}{10\sigma_{\text{GKP}}^2}\, \right], \\
    P(\eta_{p_C}, \eta_{p_T}) &= \frac{1}{\mathcal{N}_0} \exp \left[ -\frac{2\eta_{p_C}^2 + 3\eta_{p_T}^2 + 2\eta_{p_C}\eta_{p_T}}{10\sigma_{\text{GKP}}^2} \right],
    \label{eq:gkp_correlated_prob}
\end{split}
\end{equation}
where $\mathcal{N}_0 = \sqrt{20\pi^2\sigma_{\text{GKP}}^2}$, is a normalization factor. To sample CNOT gate faults, we sample eight independent shifts as shown in Fig. \ref{fig:gkp_gate} and use Eq. \ref{eq:gkp_cnot} to evaluate the effective shifts. Subsequently, we use the probability distributions above to determine the nearest valid lattice point in the phase space of the effective shifts. If this nearest point is a logical point in the GKP lattice \cite{lin2023closest, conrad2022gottesman, royer2022encoding}, we declare a gate fault.

With the above probability distributions, one can use standard techniques from probability theory to derive fault occurrence probabilities for these gate faults, similar to the single qubit cases of Eq. \ref{eq:gkp_prior} and Eq. \ref{eq:gkp_posterior}. In the interest of brevity, we shall avoid providing a detailed derivation and instead refer the interested reader to Appendix C of \cite{noh2022low} for a full derivation of these probabilities.

\section{The QLDPC Code}
\label{sec:qldpc_code}
We will now provide a detailed overview of the circuit-level noise model for the outer QLDPC code. We begin with a brief description of QLDPC codes and the syndrome measurement circuits involved in the measurement of the stabilizers of the outer QLDPC code. Subsequently, we outline the noise model for modeling faults in the syndrome measurement circuit for the QLDPC code and describe how these faults arise as a consequence of the GKP noise model developed in the previous section. Finally, we describe the decoding protocol for utilizing soft information obtained from the GKP-EC protocol to correct for circuit-level noise.

\subsection{Quantum LDPC Codes}
\label{sec:qldpc}
We consider Quantum Low Density Parity Check (QLDPC) codes belonging to the class of Calderbank-Shor-Steane (CSS) codes \cite{calderbank1996good, steane1996error}. They are discrete-variable stabilizer codes that have stabilizer generators that are separable into binary stabilizer/parity check matrices $H_X$ and $H_Z$. The respective check matrices, $H_X$ to correct Pauli $Z$ (phase flip) errors and $H_Z$ to correct Pauli $X$ (bit flip) errors, must satisfy the commuting stabilizer requirement under the binary symplectic inner product condition expressed as $H_X H_Z^{\mathsf{T}} = \mathbf{0}$, thus imposing a dual-containing property on the two classical codes. For QLDPC codes, we have the additional constraint that both $H_X$ and $H_Z$ need to be sparse matrices with both the row and column weights being bounded by a small positive integer. 

Over recent years, several constructions of QLDPC codes have been discovered with several promising properties. The hypergraph product code \cite{tillich2013quantum}, proposed by Tillich \& Z\'{e}mor, was one of the earliest QLDPC constructions that achieved a good minimum distance scaling with respect to the code length. Specifically, the hypergraph product construction allows us to construct CSS check matrices $H_X$ and $H_Z$, from two classical check matrices $H_1$ and $H_2$ in a manner that allows us to infer the encoding rate and minimum distance of the resulting quantum code as a function of the encoding rates and minimum distances of the classical codes defined by $H_1$ and $H_2$. Since the construction of good classical LDPC codes \cite{richardson2008modern, lin2021fundamentals} is a well studied topic, the hypergraph product construction allows us to build good quantum LDPC codes from good classical LDPC codes. 
\begin{figure}
	
    \centering
    \subfloat[$Z$ stabilizer measurement.\label{fig:stab_measure_z}]{%
        \scalebox{0.45}{\includegraphics[width=0.45\textwidth]{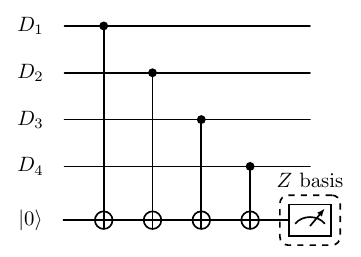}}}
    \qquad
    \subfloat[$X$ stabilizer measurement.\label{fig:stab_measure_x}]{%
        \scalebox{0.45}{\includegraphics[width=0.45\textwidth]{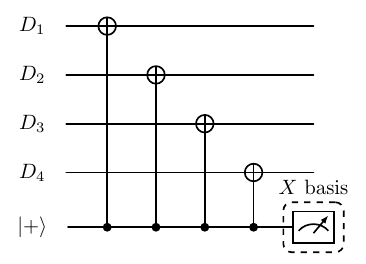}}}
    \vspace{1.5em}
    
    \caption{Syndrome measurement circuits for the measurement of a single stabilizer associated with (a) $H_X$ and (b) $H_Z$. State preparation and measurement are done in the $Z$ basis for $H_Z$ stabilizers and in the $X$ basis for $H_X$ stabilizers. The direction of the CNOT gates is also determined by the stabilizer type.}
    
    \label{fig:stab_measure}
    
\end{figure}

An important feature of hypergraph product family is that the minimum distance of this code family has a square root scaling with respect to the number of (data) qubits required. This is similar to the minimum distance scaling offered by the surface code \cite{fowler2012surface} and its variants, but the encoding rates offered by the hypergraph product construction are significantly larger. In \cite{panteleev2021quantum}, the authors exploited protograph construction methods \cite{thorpe2003low} from classical coding theory to build a variant of the hypergraph product family, known as the lifted product family. Certain classes of lifted product product codes have been shown to have a linear minimum distance scaling with respect to the block length, which combined with the high encoding rates already provided by the hypergraph product construction, have the potential to significantly reduce the overhead requirements for achieving fault tolerant quantum error correction.

Another important class of QLDPC codes is the class of bicycle bivariate codes introduced in \cite{bravyi2024high}. While the minimum distance scaling of this code family is not as high as that of the lifted product family, these codes have several advantages for nearer term implementations in the context of having a lower number of long range connections, which offers a more hardware friendly approach. For this work, we will investigate the performance of two QLDPC codes, one from the lifted product family and the other from the bicycle bivariate family under the circuit-level noise model.

We will assume that the decoding and correction of both types of errors are done separately using a separate decoder for each type of error. For measuring the syndromes that will be fed into these decoders, we make use of the circuits of the form shown in Fig. \ref{fig:stab_measure}. Specifically, each row in $H_X$ and $H_Z$ corresponds to a separate stabilizer. For each stabilizer in $H_Z$, we use a circuit of the form shown in Fig. \ref{fig:stab_measure_z}, where $D_1, D_2, D_3, D_4$ represent the data qubits involved in a given stabilizer. For each $Z$ stabilizer, a separate ancilla qubit is used, which is prepared in the $\ket{0}$ state and measured in the $Z$ basis. Similarly, for $X$ stabilizers, we use circuits of the form shown in Fig. \ref{fig:stab_measure_x}, where each ancilla is prepared in the $\ket{+}$ state and measured in the $X$ basis.
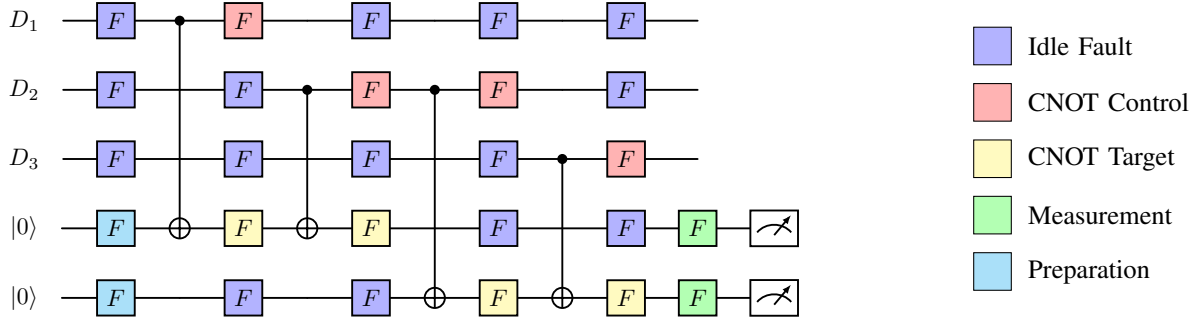
\begin{figure*}
    
    \centering
    
    \tikzset{
        ctrl/.style={fill=red!30},
        idle/.style={fill=blue!30},
        meas/.style={fill=green!30},
        targ/.style={fill=yellow!30},
        prep/.style={fill=cyan!30}
    }
    
    \vspace{0.5em}
    \begin{minipage}[c]{0.75\linewidth}
        \centering
        \scalebox{1.8}{\includegraphics[width=0.45\textwidth]{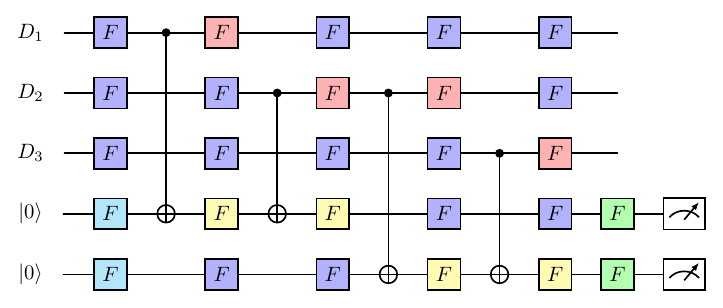}}
    \end{minipage}%
    \hfill
    \begin{minipage}[c]{0.25\linewidth}
        \centering
        \scalebox{1}{\begin{tikzpicture}
            \draw[fill=blue!30] (0,0) rectangle (0.5,0.5);
            \node[right] at (0.6,0.25) {Idle Fault};
            
            \draw[fill=red!30] (0,-0.75) rectangle (0.5,-0.25);
            \node[right] at (0.6,-0.5) {CNOT Control};

            \draw[fill=yellow!30] (0,-1.5) rectangle (0.5,-1);
            \node[right] at (0.6,-1.25) {CNOT Target};

            \draw[fill=green!30] (0,-2.25) rectangle (0.5,-1.75);
            \node[right] at (0.6,-2) {Measurement};

            \draw[fill=cyan!30] (0,-3) rectangle (0.5,-2.5);
            \node[right] at (0.6,-2.75) {Preparation};
        \end{tikzpicture}}
    \end{minipage}
    \vspace{1.5em}
    
    \caption{Circuit noise model for the 3-qubit bit-flip repetition code. The boxes marked $F$ indicate potential faults with $F \in \{I, X\}$. The color of the box indicates the type of fault as recorded on the right side of the figure. Each qubit is a GKP qubit and is supported by two ancilla GKP qubits (not shown) to perform the GKP error correction protocol. The noisy GKP protocol is assumed to be performed at each fault location so that analog-valued information about these circuit faults can be obtained. We will use $r_Z$ and $r_X$ to denote the number of fault locations (in this case $r_Z = 27$) per measurement round of $H_Z$ and $H_X$ respectively.}
    
    \label{fig:rep_circuit}
    
\end{figure*}

\subsection{Circuit Level Noise}
\label{sec:circuit_noise}
We will use the circuit-level noise model to model the effect of circuit faults in the syndrome measurement and decoding of the outer code in a concatenated QLDPC-GKP scheme. An illustration of the circuit-level noise model for a 3-qubit bit-flip repetition code \cite{gottesman2016surviving} is shown in Fig. \ref{fig:rep_circuit}. We assume that the CNOT gates are performed in discrete time-steps, which we will refer to as gate levels. In Fig. \ref{fig:rep_circuit}, we have shown only one gate per level for the sake of clarity, but in general, several gates may be performed in parallel within the same gate level if they operate on disjoint sets of qubits.

Each gate level is followed by a layer of fault operators, which represent circuit faults that may occur with a certain probability. In Fig. \ref{fig:rep_circuit}, we only consider faults of type $X$ since the bit-flip code only corrects for $X$ errors. In general, faults of $Z$ type will also be present. Five different types of faults can be seen in Fig. \ref{fig:rep_circuit}, which occur based on their location in the circuit. These are:
\begin{itemize}
    \item[(1)] \emph{Idling faults}: These occur on a given qubit if the qubit was not acted upon by a CNOT gate in the preceding gate level.
    \item[(2)] \emph{Control faults}: These occur on a given qubit if the qubit was the control of a CNOT gate in the preceding gate level.
    \item[(3)] \emph{Target faults}: These occur on a given qubit if the qubit was the target of a CNOT gate in the preceding gate level.
    \item[(4)] \emph{Preparation faults}: These occur on the ancilla qubits at the beginning of the circuit.
    \item[(5)] \emph{Measurement faults}: These occur on the ancilla qubits at the end of the circuit, just prior to measurement.
\end{itemize}

In general, each type of fault will have a different noise model and will be characterized by a different fault occurrence probability. A quick glance at Fig. \ref{fig:gkp_faults} reveals that the above types of faults are in one-to-one correspondence with the four scenarios discussed in Sec. \ref{sec:gkp_noise}. We will assume that each qubit in Fig. \ref{fig:rep_circuit} is a GKP qubit and is also supported by two additional GKP ancilla qubits to perform the teleportation-based GKP-EC protocol. Hence, the circuit of Fig. \ref{fig:rep_circuit} requires a total of 15 oscillator modes. Furthermore, we will assume that a round of the noisy GKP protocol is performed at each of the fault locations shown in Fig. \ref{fig:rep_circuit}. Hence each fault block, marked $F$ in Fig. \ref{fig:rep_circuit} attempts to perform a round of GKP-EC to correct for any faults that occur at the fault location. Moreover, each fault box also produces soft information in the form of Eq. \ref{eq:gkp_posterior} that can be used to inform the decoding process of the outer decoder.

\begin{remark*}
    Note that the number of idling faults and the locations of idling and gate faults will depend strongly on the choice of measurement schedule used for the syndrome measurements. In Sec. \ref{sec:lp_sim}, we will investigate the effect of using measurement schedules with varying depths and demonstrate that the use of an optimized depth measurement schedule is critical for properly utilizing soft information from the inner code in the outer decoder.
\end{remark*}

\subsection{Decoding Circuit-Level Noise}
\label{sec:decoder}
We will now outline the decoding process for correcting faults induced through the noisy GKP fault mechanisms described in Sec \ref{sec:gkp_noise}. First, we will show how the circuit-level parity check matrix for a given QLDPC code is obtained. We will then use this circuit-level parity check matrix in conjunction with the Belief Propagation decoder with Ordered Statistics Decoding (BP-OSD) \cite{panteleev2021degenerate} to decode faults in the circuit-level noise model. Soft information obtained from the GKP measurements, as described in the previous section, will be used to initialize the log-likelihood values of this decoder.

We will assume three rounds of repeated measurements of each parity check matrix ($H_X$ and $H_Z$) to allow the decoder to correct for potential syndrome errors. As mentioned at the end of Sec. \ref{sec:qldpc}, we assume that each type of error ($X$ or $Z$) is decoded separately. Hence, we assume that the measurement order involves 3 rounds of $X$ measurements followed by 3 rounds of $Z$ measurements and so on. Henceforth, we shall limit our focus to the correction of $X$ errors (using the check matrix $H_Z$) and assume that the decoding of $Z$ errors occurs similarly with a comparable error rate. The measurement of the parity check matrix $H_Z$ will be done via a circuit similar to that shown in Fig. \ref{fig:rep_circuit}. Let us assume that $H_Z$ is an $m \times n$ matrix and that the measurement of $H_Z$ and $H_X$ involve a total of $r_Z$ and $r_X$ fault locations (see Fig. \ref{fig:rep_circuit}) respectively. Then, the circuit-level parity check matrix $H_Z^{circ}$ will be a binary matrix of dimensions $3m \times 3(r_X + r_Z)$. The rows correspond to the syndromes obtained from 3 repeated rounds of measuring $H_Z$ and the each column corresponds to a separate fault. The entry at the $i^{th}$ row and $j^{th}$ column will be a one if the corresponding fault triggers the corresponding syndrome, and zero otherwise. The set of syndromes triggered by a given fault can easily be obtained via a stabilizer tableau simulation \cite{gidney2021stim}. Note that we consider faults (of type $X$) in the syndrome measurement circuits of both $H_X$ and $H_Z$ but only the syndromes of $H_Z$, since faults of type $X$ only trigger $Z$ syndromes.

In principle, the parity check matrix $H_Z^{circ}$ may now directly be used with a BP-OSD decoder to correct for faults in the circuit-level noise model. However, for any reasonably sized QLDPC code, the number of fault locations $r_X + r_Z$ will likely be prohibitively large. Hence, the matrix $H_Z^{circ}$ will need to be simplified by removing redundant columns before it can be used for decoding. To do this, note that the measurement circuits for both $H_Z$ and $H_X$ will likely contain several equivalent faults that yield identical syndromes and have an equivalent effect on the data qubits. As an example, consider the first 4 idling faults on $D_3$ in Fig. \ref{fig:rep_circuit}. Such faults will correspond to identical columns in $H_Z^{circ}$ and hence, all of these 4 identical columns can be replaced by a single column. This optimization significantly reduces the size of $H_Z^{circ}$.

To account for the fact that this single column now represents a set of four faults, we will also need to update the fault occurrence probabilities of these faults. Specifically, consider a set of faults $F_1, F_2, \dots, F_N$ being represented by a single fault $F$. Let $P_1, P_2, \dots, P_N$ be the fault occurrence probabilities of the faults $F_1, F_2, \dots, F_N$, respectively, and let $P$ be the fault occurrence probability of $F$. The fault $F$ is said to occur if an odd number of the faults in $F_1, F_2, \dots, F_N$ occur, which happens with probability,
\begin{equation}
    P = \frac{1}{2} - \frac{1}{2}\prod_{i=1}^n (1-2P_i).
    \label{eq:fault_merge}
\end{equation}
Thus, we summarize the decoding process as follows. Given parity check matrices $H_X$ and $H_Z$ along with syndrome measurement circuits of the form of Fig. \ref{fig:rep_circuit} for each of these two matrices, we evaluate the circuit-level parity check matrix $H_Z^{circ}$ using the methods described above. We then remove redundant columns from $H_Z^{circ}$ and record the set of fault locations that are represented by each column in $H_Z^{circ}$. For a given decoding run, we then sample faults using the methods described in Sec. \ref{sec:gkp_noise}. Subsequently, we evaluate the syndrome corresponding to these sampled faults using $H_Z^{circ}$ and run a BP-OSD decoder using the check matrix $H_Z^{circ}$ to benchmark decoder performance. We will benchmark decoder performance in the following three cases:
\begin{itemize}
    \item[(1)] \emph{Case 1: No soft information.} \\ 
    In this case, the BP-OSD decoder is initialized with identical log-likelihood ratios (LLR) for each column of $H_Z^{circ}$. Hence, the decoder has no information regarding the fault occurrence probabilities of the various columns.
    \item[(2)] \emph{Case 2: Precomputed Reliability.} \\
    In this case, we evaluate the fault occurrence probabilities using Eq. \ref{eq:gkp_prior} and feed it to the BP-OSD decoder as initial LLR values for each column. For columns representing multiple faults, we combine these fault occurrence probabilities using Eq. \ref{eq:fault_merge}. In this case, the decoder has some information about the relative fault occurrence probabilities of each column but still has no information about the sampled faults.
    \item[(3)] \emph{Case 3: Real Time Soft Information.} \\
    Here, we use the sampled GKP measurement $\eta$ obtained from the fault blocks of Fig. \ref{fig:rep_circuit} to evaluate the fault occurrence probabilities using Eq. \ref{eq:gkp_posterior}. Similar to the above case, we use Eq. \ref{eq:fault_merge} to combine fault occurrence probabilities for columns representing multiple faults. In this case, the decoder is provided with a maximal amount of soft information regarding the fault sample.
\end{itemize}
This completes the description of the decoding process. In the next section, we will benchmark the above three scenarios using a memory experiment on two different QLDPC codes, one from the bicycle bivariate family and the other from the lifted product family.

\begin{remark*}
    In the previous sections, we have described how soft information from the GKP measurements can be obtained in the form of probability values in the form of Eq. \ref{eq:gkp_prior} and Eq. \ref{eq:gkp_posterior}. For the actual experiments (to be described in Sec. \ref{sec:results}), involving the BP-OSD decoder, it is prudent to use these values in the log-likelihood domain rather than the probability domain, in order to ensure numerical stability. For a given probability of error $P$ the log-likelihood value to be used in the decoder is given by,
    \begin{equation}
        L = \ln\left(\frac{1-P}{P}\right).
        \label{eq:llr}
    \end{equation}
\end{remark*}

\section{Numerical Results}
\label{sec:results}
We will now describe the results of a quantum memory experiment performed on a concatenated QLDPC-GKP architecture with the noise model and decoding procedure detailed in the previous section. We begin with a description of the simulation setup and the quantum memory experiment that we use to benchmark decoding performance. Subsequently, we proceed to discuss the results of our experiments on two classes of quantum LDPC codes, namely, the bicycle bivariate code and the lifted product code. In the latter case, we also discuss the effect of using different measurement schedules.

\subsection{Simulation setup}
\label{sec:sim_setup}
To benchmark the performance of the circuit-level decoder in the presence of GKP soft information, we perform a set of quantum memory experiments on a concatenated QLDPC-GKP architecture with the QLDPC code being chosen as detailed above. Specifically, we use the following simple procedure to benchmark decoder performance. First, we sample a set of faults $\vec{f}_0$ according to the noise model described in Sec. \ref{sec:gkp_noise}. This set of faults corresponds to an effective error $\vec{e}_0$ on the data qubits, which can be obtained via stabilizer tableau simulation \cite{gidney2021stim}. We then use the decoding procedure of Sec. \ref{sec:decoder} to get an estimate $\vec{f}$ of the fault locations. Given this estimate on the set of faults, we evaluate the effective error $\vec{e}$ on the data qubits that occur due to these faults via a stabilizer tableau simulation \cite{gidney2021stim}, and apply the corresponding recovery operation. Having applied the recovery operation, we evaluate the post-recovery effective error on the data qubits as $\vec{e'} = \vec{e}_0 + \vec{e}$. Ideally, we would like to have $\vec{e'} = 0$, but this is likely not going to be true, even in the case of successful decoder operation. To see why this is so, consider for example, the last idling fault in the qubit $D_1$ in Fig. \ref{fig:rep_circuit}. This fault will not trigger any of the syndromes and hence, it is impossible for the decoder to correct for this fault. Consequently, this fault will have to be dealt with in the next decoding cycle. We say that the decoder is successful if the error $\vec{e'}$ is within the error correcting capability of the base code $H_Z$. Hence, to benchmark the decoder, we can simply perform a destructive measurement of the data qubits (in the $X$ basis for benchmarking $H_Z$) and then feed it into another decoder equipped with the matrix $H_Z$. If this decoder succeeds in correcting the error $\vec{e'}$, we declare a decoding success for the overall experiment sample. For the full experiment, we repeat this process in a Monte Carlo sense until at least $1000$ decoder failures are observed and evaluate the frame error rate (FER) as the ratio of the number of failures (1000) to the number of Monte Carlo trials.
\begin{figure}
    \centering
        \includegraphics[width=0.48\textwidth]{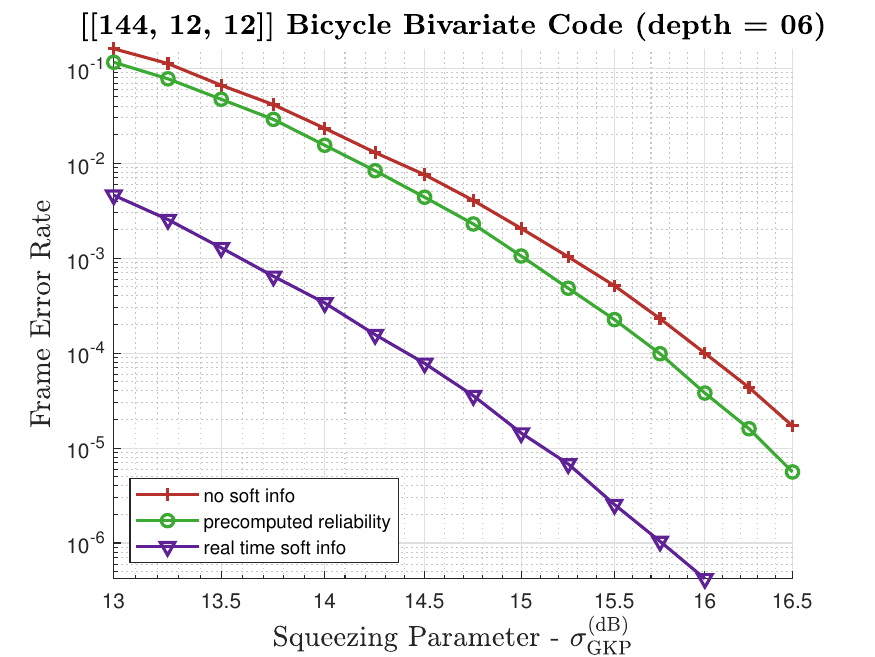}
    \caption{Monte Carlo simulation results for the quantum memory experiment performed on a concatenated QLDPC-GKP scheme with a $[[144, 12, 12]]$ bicycle bivariate QLDPC code, with a measurement schedule of depth 6.}
    \label{fig:bb_sim}
\end{figure}

\subsection{Bicycle bivariate code}
\label{sec:bb_sim}
We now present the results of the quantum memory experiment described above for a bicycle bivariate code. For our simulations, we have used a bicycle bivariate code with parameters $[[144, 12, 12]]$ from \cite{bravyi2024high}. An optimized measurement schedule for the syndrome measurement for this code was also provided in \cite{bravyi2024high} and we use this same schedule with a faulty circuit similar to Fig. \ref{fig:rep_circuit} for our experiment. This measurement schedule has a total CNOT gate depth of $6$ per stabilizer type ($X$ or $Z$), yielding a total depth of $12$ per measurement round. Subsequently, we build the circuit-level parity check matrix $H_Z^{circ}$ of this code using the methods described in Sec. \ref{sec:decoder}. This circuit-level matrix has dimensions $216 \times 3600$.

The results of our experiments are shown in Fig. \ref{fig:bb_sim}. The experiment is performed for each of the three cases of soft information availability discussed in Sec. \ref{sec:decoder}. As expected, we have a demonstrable improvement in the frame error rate in the presence of soft information. We observe that there is a small improvement obtained by using precomputed reliabilities (as per Eq. \ref{eq:gkp_prior}), when compared to the case of uniform LLR initialization for the decoder. But a much more significant improvement is obtained through the use of real-time soft information (as per Eq. \ref{eq:gkp_posterior}) obtained from analog GKP stabilizer measurements from the fault blocks of Fig. \ref{fig:rep_circuit}. Indeed, the use of analog GKP stabilizer measurements for decoding at the outer layer improves the error rates for this particular code by up to two orders of magnitude. This implies that the real time soft information obtained from sample-to-sample execution of the GKP-EC blocks is indeed critical for the outer decoder to be able to accurately infer fault locations.

\begin{remark*}
    The optimal schedule for the bicycle bivariate code introduced in \cite{bravyi2024high} involves simultaneous measurements of both the $X$ and $Z$ stabilizers. For this work, we have a slightly modified version, wherein we split up the stabilizer measurements of different types to occur one after the other without overlap among themselves. Strictly speaking, this is suboptimal with respect to measuring both stabilizer types simultaneously but will suffice for the purposes of demonstrating the improvements offered by the use of soft information. In Sec. \ref{sec:lp_sim}, we show that lowering the depth of the measurement circuit significantly improves the reliability of the GKP soft information and hence, it is likely that using a measurement schedule with interleaved $X$ and $Z$ measurements will further improve the error rates of Fig. \ref{fig:bb_sim}.
\end{remark*}
\begin{figure*}
    \centering
    \subfloat[FER with a measurement schedule of depth 40.\label{fig:lp_sim_40}]{%
        \centering
        \includegraphics[width=0.48\textwidth]{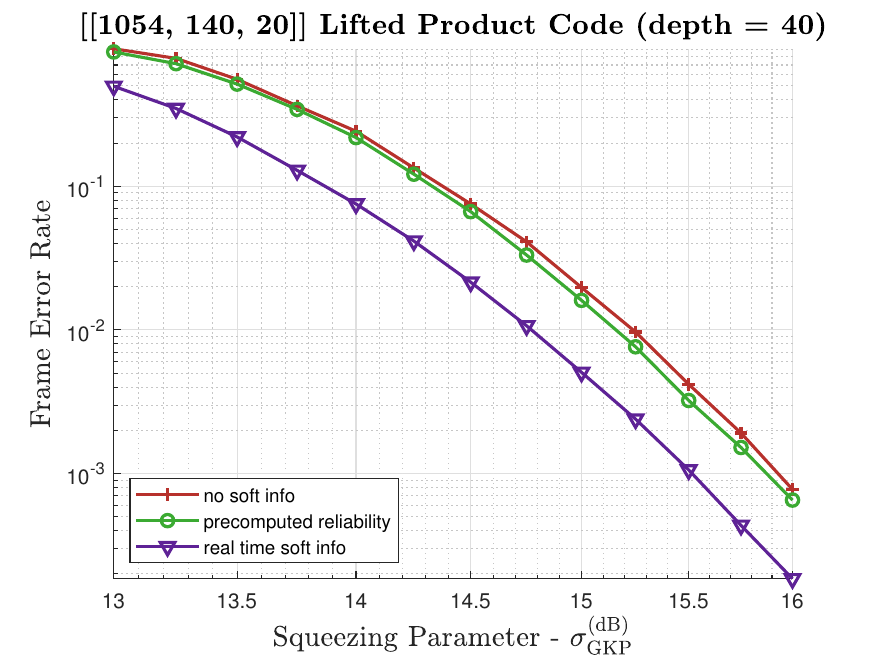}}
    \hfill
    \subfloat[FER with a measurement schedule of depth 8.\label{fig:lp_sim_08}]{%
        \centering
        \includegraphics[width=0.48\textwidth]{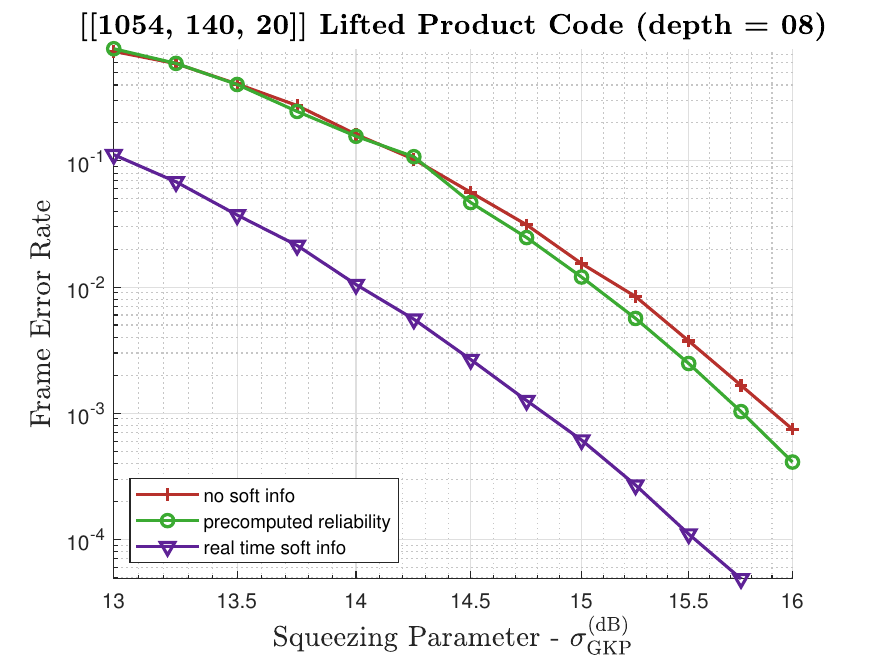}}
    \caption{Monte Carlo simulation results for the quantum memory experiment performed on a concatenated QLDPC-GKP scheme with a $[[1054, 140, 20]]$ lifted product QLDPC code, with (a) a measurement schedule of depth 40 and, (b) a measurement schedule of depth 8.}
    \label{fig:lp_sim}
\end{figure*}

\subsection{Lifted product code}
\label{sec:lp_sim}
Next, we investigate the performance of our decoding approach for a QLDPC code from the lifted product family. For this code, we choose two identical base matrices of size $3 \times 5$ and apply the lifted product construction \cite{panteleev2021quantum} with a lift size of $31$ to get check matrices $H_X, H_Z$ of dimensions $465 \times 1054$. The resulting code has parameters $[[1054, 140, 20]]$, where we have used the standard notation \cite{gottesman2016surviving} of $[[n, k, d]]$ to represent a quantum code with $n$ physical qubits, $k$ logical qubits and a minimum distance of $d$. The corresponding circuit-level parity check matrix $H_Z^{circ}$ has dimensions $1395 \times 30132$. While a general approach to depth optimization for the syndrome measurement circuit for a lifted product code is not known, brute force optimization techniques can be applied for specific cases. In our case, we investigate the performance for two different measurement schedules, one of depth $40$ (per stabilizer type per measurement round) and the other with an optimal depth of $8$ (per stabilizer type per measurement round).

The results of the experiment are shown in Fig. \ref{fig:lp_sim}. As in the case of the bicycle bivariate code, we observe that significant gains are observed only when real-time soft information from the inner GKP code is passed to the outer decoder. We also observe that optimizing the measurement depth plays a critical role in ensuring the reliability of this soft information. As observed in Fig. \ref{fig:lp_sim}, the gain due to soft information for the case of the schedule with depth 40 is approximately half an order of magnitude, whereas the corresponding gain for the optimal schedule with depth 8 is approximately one order of magnitude in the frame error rate. To understand why this is the case, we recall the faulty circuit of Fig. \ref{fig:rep_circuit}. Note that the number of preparation and ancilla faults will be equal to the number of ancilla qubits and the number of control and target faults will be equal to the number of CNOT gates (or equivalently, the number of ones in $H_Z$). But the number of idling faults increases with an increase in the depth of the measurement circuit.

To understand the effect of a larger number of idling faults, consider a series of idling faults $F_1, F_2, \dots F_N$ each occurring with a probability $p_i$. As described in Sec. \ref{sec:decoder}, this series of faults is equivalent to a single idling fault with a fault occurrence probability $p$ given by Eq. \ref{eq:fault_merge}. At the end of Sec. \ref{sec:decoder}, we noted that the actual computations of the decoder are carried out in the log-likelihood domain rather than the probability domain for reasons relating to numerical stability. Transforming Eq. \ref{eq:fault_merge} to the log-likelihood domain gives us the well known tanh rule and its min-sum approximation,
\begin{equation}
\begin{split}
    L &= 2\,\text{arctanh}\left(\prod_{i=1}^N \tanh{\frac{L_i}{2}}\right) \\
    &\approx \prod_{i=1}^N \left(\text{sign}(L_i)\right) \cdot \min|L_i|,
\end{split}
\end{equation}
where $L$ and $L_i$ are related to $p$ and $p_i$ respectively through Eq. \ref{eq:llr}. The magnitude of the log-likelihood value is a measure of the reliability of the corresponding fault information \cite{richardson2008modern} and therefore, the above equation simply tells us that the reliability of soft information obtained from a chain of faults is equal to the minimum reliability of the soft information from each fault in the chain. Hence, a longer measurement schedule implies a lower reliability of soft information due to longer chains of idling faults. This effect can be clearly seen in our experimental results by simply comparing the gains offered by the use of soft information in Fig. \ref{fig:lp_sim_40} and in Fig. \ref{fig:lp_sim_08}.

\section{Conclusions}
\label{sec:conclusion}
In this paper, we have investigated the performance of concatenated QLDPC-GKP architectures under the circuit-level noise model. Specifically, we assume that a round of the teleportation-based protocol for GKP-EC is performed after every gate level during the syndrome measurement process for the outer QLDPC code. These GKP measurements provide us with additional analog soft information regarding the fault values which then can be used by a BP-OSD decoder operating on the circuit-level parity check matrix of the code for improving frame error rates. For benchmarking purposes, we have considered three scenarios, namely, no availability of soft information to the decoder, feeding precomputed fault reliability information into the decoder and feeding circuit-level soft information (from GKP stabilizer measurements) to the decoder. Experiments indicate significant improvements only in the third case, indicating that real-time soft information obtained from GKP syndrome measurements plays a crucial role in enabling the outer decoder to correctly account for the effects of circuit faults. We also investigate the effect of varying the depth of the measurement schedule on the reliability of soft information, wherein our experiments indicate that optimizing the depth of the measurement schedule is critical for preserving the reliability of soft information obtained from the inner GKP code.

While preliminary results on the use of circuit-level soft information for QLDPC-GKP decoding are promising, there are several challenges that will need to be addressed in future research. For instance, we have also assumed that a round of GKP error correction always follows each gate level in the outer code syndrome measurement circuit. In general, this need not be the case and one could, in principle, think of designing a more strategic placement of these GKP blocks to specifically target faults that lead to large correlated errors in the data qubits. Finally, we remark that the sizes of the circuit-level parity check matrices that have been obtained are still relatively large to admit any kind of practical implementation at this stage, which implies that the development of low-complexity circuit-level decoders is an important challenge that will need to be addressed in future research.

\section*{Acknowledgment}
SKB and BV are funded by the US Department of Energy, Office of Science, National Quantum Information Science Research Centers, Superconducting Quantum Materials and Systems Center (SQMS) under contract No. DE-AC02-07CH11359. AKP, NR and MP are funded in part by the NSF under grants CIF-2420424 and CIF-2106189.

% \IEEEtriggeratref{20}
\bibliographystyle{IEEEtran}
\bibliography{IEEEabrv, bib/refs}

\end{document}